# *Quantum Mechanics Unscrambled*[*]

*Jean-Michel Delhôtel*[1][†]

**Abstract.** *Is quantum mechanics about 'states'? Or is it basically another kind of probability theory? It is argued that the elementary formalism of quantum mechanics operates as a well-justified alternative to 'classical' instantiations of a probability calculus. Its providing a general framework for prediction accounts for its distinctive traits, which one should be careful not to mistake for reflections of any strange ontology. The suggestion is also made that quantum theory unwittingly emerged, in Schrödinger's formulation, as a 'lossy' by-product of a quantum-mechanical variant of the Hamilton-Jacobi equation. As it turns out, the effectiveness of quantum theory qua predictive algorithm makes up for the computational impracticability of that master equation.*

"Our present quantum mechanical formalism is a peculiar mixture describing in part laws of Nature, in part incomplete human information about Nature – all scrambled up together by Bohr into an omelette that nobody has seen how to unscramble. Yet we think the unscrambling is a prerequisite for any further advance in basic physical theory.

…if quantum theory were not successful pragmatically, we would have no interest in its interpretation. It is precisely because of the enormous success of the QM mathematical formalism that it becomes crucially important to learn what that mathematics means. To find a rational physical interpretation of the QM formalism ought to be considered the top priority research problem of theoretical physics; until this is accomplished, all other theoretical results can only be provisional and temporary."

– E.T. Jaynes

---

[*] This paper is the final (v3) version of a 2004 preprint, to the original title of which it reverts. The previous two Arxiv.org postings should henceforth be ignored, with the understanding that no further update is intended. For those readers who missed the hint, the title makes slightly ironical reference to Daniel Dennett's *Consciousness Explained*. Needless to say, this implies no pretence of providing any 'final' answer to some vexing questions, but should be regarded as an invitation to tackle those without smugness and prejudice.
[†] JMRL.Delhotel@alumni.lse.ac.uk

## 1. Quantum mechanics without 'states'?

> "The best response so far is from Pauli, who at least admits that the use of the word 'state' [*Zustand*] for the psi-function is quite disreputable[2]." – *E. Schrödinger*

Should we follow Pauli and refrain from thinking of Hilbert space vectors or 'wave functions' as representatives of *quantum states* – given what the word 'state' is intended to mean: *how something is* and, by implication, what makes it such (which properties it *has* etc.) ? Given the lack of a satisfactory, widely agreed upon 'interpretation' of the basic formalism of quantum theory – some eight decades after its inception – this cannot be lightly dismissed as an idle question. Back in the late 1920s, Niels Bohr pointed out that, whilst observation implies "interactions with suitable agencies of measurement, not belonging to the system", it is also the case that "the definition of the state of a physical system, as ordinarily understood, claims the elimination of all external disturbances[3]." In atomic physics, however, interactions implied by the operation of probing devices can never be assumed to have a negligible effect on our descriptive and predictive abilities. As a result, the idea of an individual system 'having a state' or its 'being in' one can no longer, in general, be consistently maintained: "an unambiguous definition of the state of a system is naturally no longer possible, and there can be no question of causality in the ordinary sense of the word." Nothing in the quantum-mechanical formalism would qualify as a meaningful representative of the state of a physical system. Whoever, by contrast, keeps regarding a Hilbert space vector as one such representative would overlook a key lesson of the quantum 'revolution', mistaking its epistemic and pragmatic implications for manifestations of a baffling ontology: one according to which quantumstuff could at once be – no kidding – 'something like' this or that, or perhaps this and that, but neither this nor that or partly this and partly that.

Throughout his discussions with Bohr and until his last breath, Albert Einstein remained convinced that quantum mechanics is irrevocably inadequate as a truly fundamental physical theory, since it does not fulfil the descriptive and explanatory aims which, he believed, are those of such a theory. Einstein was adamant that *fundamental* physical laws cannot "consist in relations between *probabilities* for the real things, but [should consist in] relations concerning the things themselves[4]." As he saw it, the probabilistic use of 'state vectors', or wave functions, would rather indicate that they refer to *ensembles*. Whichever way Einstein conceived[5] of such ensembles, he persisted in regarding the 'mechanics' of Heisenberg or Schrödinger as a makeshift, whilst keeping up hopes that it would, sooner or later, give way to a truly worthy account of the interactions and properties of microphysical systems.

---
[2] Erwin Schrödinger, referring to the EPR challenge (Einstein *et al.* 1935) in a letter to Albert Einstein dated 13 July 1935; quoted by Fine 1986, p.74.
[3] Bohr 1928; quoted from Jammer 1989, p.366.
[4] Einstein, quoted by Fine 1986, pp.100-101.
[5] Deltete and Guy 1990.

Bohr's and Einstein's attitudes sharply contrast with those of some of their most prominent colleagues, and in particular Dirac's. Indeed, the entire introduction to his *Principles of Quantum Mechanics* is devoted to quantum states and to their representation[6]. Dirac regards the task of setting up a novel kind of mechanics as a necessary response to the recognition of "a limit to the gentleness with which observations can be made[7]" on physical systems. The *principle of superposition* would be "one of the most fundamental and drastic[8]" of a new set of "laws of nature", which it would be necessary to introduce because such disturbance cannot be eliminated. That much sounds close enough to Bohr's own views. However, Dirac also maintains that, as "a mathematical procedure[9]", expressing a *state* as a superposition of various other *states* "is always permissible, independent of any reference to physical conditions, like the procedure of resolving a wave into Fourier components.[10]" Permissible as such an expression may be, one should be wary of interpreting relationships between linearly superposed *states*, since those relationships "cannot be explained in terms of familiar physical concepts[11]." Dirac is well aware of the difficulties that a state-based account of quantum mechanics is bound to raise: we certainly "cannot in the classical sense picture a system *being* partly *in* each of two states and see the equivalence of this to the system being completely in some other state. There is an entirely new idea involved, to which one must get accustomed and in terms of which one must proceed to build up an exact mathematical theory, without having any detailed classical picture[12]." By way of addressing the "new idea", Dirac comes close to venturing an ontological reading of superpositions: "When a state is formed by the superposition of two other states, it will *have properties* that *are* in some vague way [!] *intermediate between* those of the two original states and that *approach* [?] *more or less closely to those of either of them* according to the greater or less 'weight' attached to this state in the superposition process[13]." Aware that it is all getting shaky, Dirac hastens to add: "the intermediate character of the state formed by superposition thus expresses itself through the *probability* of a particular *result* for an observation being intermediate between the corresponding probabilities for the original states, *not* through the result itself being intermediate between the corresponding results for the original states[14]." But there remains the thornier question of hypothetically possessed properties in relation to their presumed numerical, empirically accessed indicators (the measurement results). Does it make any sense to think of such properties as in some way 'intermediate' between observationally ascertainable attributes – if the latter are what eigenvalues denote? Dirac brushes aside questions that one may be tempted to ask regarding the definiteness of 'possessed values' between observations, regarding them as 'metaphysical' issues of no concern to the physicist[15]. The matter is settled: the principle of superposition derives both its necessity and its effectiveness from the recognition of an irreducible susceptibility of microphysical objects to disturbance, which just "demand[s] indeterminacy in the results of observation[16]." Once we have come to terms with the irreducible character of that susceptibility, developing a probabilistic framework with inherent linear features – superposition obliges – should not give rise to any qualms.

---

[6] Dirac 1958.
[7] Dirac 1958, p.4.
[8] *Ibid.*
[9] *Ibid.*, p.12.
[10] *Ibid.*
[11] *Ibid.*
[12] *Ibid.*(italics added).
[13] *Ibid.*, p.13 (italics added).
[14] *Ibid.* The whole sentence appears in italics in Dirac's text.

A view of quantum mechanics as fundamentally about the specification, evolution and structural changes of 'quantum states' crystallized with the first abstract formulations of the principles of the new theory. After Dirac, von Neumann's influential monograph[17] greatly contributed to promoting a conception of quantum mechanics where 'quantum states' are pervasive (it is no coincidence that von Neumann should have been led, in that very same endeavor, to confronting a 'measurement problem' he could not solve[18]). However, in spite of Dirac's claims, an idea of *quantum state* was certainly not forced upon the physicists as a direct response to theoretical puzzles, e.g. the black-body problem or the discrete patterns of atomic spectra. Heisenberg's initial matrix formalism ignored such states, the very idea of which was fundamentally at odds with his focus on physical, experimentally accessible quantities and the way they evolved. Nonetheless, Dirac's and von Neumann's authority prevailed, and we have been stuck with those 'quantum states' ever since. Like it or not, thinking habits and word usage are intertwined and not so easily disentangled. Thus, referring to *states*, however loosely (or especially so), can breed some ill-advised expectations. One can therefore do worse – much worse – than be mindful of the critical attitude of some, at least, of the founding fathers; for they may not, after all, have been the elderly holdouts they have often been portrayed as.

As we shall see in Section 2, a host of variously convincing, but certainly significant derivations of the elementary formalism of quantum theory converge towards the conclusion that it functions as a specific kind of linear predictive scheme, which differs in some essential respects from ordinary ('classical') ways of evaluating and computing probabilities. All those derivations have the common characteristic of proceeding from a minimal set of assumptions that are essentially neutral when it comes the 'quantal', microscopic or even physical nature of the 'systems' the scheme is applied to. Indeed, that scheme may well be relevant to conceptual domains and areas of experience which have little, if anything, to do with the subject matter of physical science. 'State' assignments would then be just a way of encapsulating a compendium of probabilities of possible outcomes ('events' or measurement results), any such assignment being referred to what will be, lacking a better word, called a *preparation*.

---

[15] In a letter to Einstein (dated 18 November 1950), Schrödinger expresses his dissatisfaction with Dirac's and others' use of probability, which in his view just covers up basic problems or inadequacies: "It seems to me that the concept of probability is terribly mishandled these days. Probability surely has as its substance a statement as to whether something *is* or *is not* the case – an uncertain statement, to be sure. But nevertheless it has meaning only if one is indeed convinced that the something in question quite definitely either *is* or *is not* the case. A probabilistic assertion presupposes the full reality of its subject." (Przibram 1967, p.37).

[16] Dirac 1958, p.14.

[17] Von Neumann 1932.

[18] And that no one can claim to have solved yet – unsurprisingly enough, if the problem does not so much need to be solved as to be dissolved.

The word 'preparation' should not be taken too literally, e.g. as implying operations that would, or could, be consciously performed on a system. The assignment might just as well be the result of an informed guess at the 'predictive potential' of a given experimental set-up or observational situation. In any case, the resulting 'state' would have no objective significance besides its predictive function. Making use of such a construct would require (i) specifying how it transforms with time; and (ii) an effective algorithm for computing outcome probabilities given its initial or time-evolved expression. Such specification and rules are precisely what quantum theory provides.

Those realists who believe that no physical theory is worth its salt unless it delivers objective truths about a pre-existing and pre-structured reality 'out there' will certainly recoil at the above characterization of quantum mechanics. However, acknowledging the chief predictive aim of the theory, and that its mathematical structure is wholly determined by the basic requirements such a purpose entails, need not amount to one's surrendering to some barren form of operationalism. Rather, recognizing its fundamentally predictive role is a necessary first step towards a positive reassessment of the nature and scope of the standard quantum-mechanical formalism. As it will be argued in Section 4, a view of quantum theory as a linear predictive scheme can be consistently maintained together with a principle-based, if tentative account of the occurrence of and the need for *quantization*. Quantum theory, qua linear predictive scheme, will be seen to emerge as a result of selecting, either unwittingly (as it seems to have been the case in the historical development of quantum mechanics) or in full awareness of what such selection presupposes and implies, a certain class of solutions to Schrödinger's equation (itself the outcome of linearizing a dynamical 'master' equation). Valuable light may thus be shed on vexed questions, such as whether standard quantum mechanics can be said to provide a complete picture of dynamics in the microrealm, or whether resorting to probability is indispensable.

## 2. Quantum theory as a predictive scheme

> Erwin with his psi can do
> Calculations quite a few.
> But one thing has not been seen
> Just what psi really mean.
> – F. Bloch

In the early 1940s Jean-Louis Destouches, a physicist with philosophical inclinations, made a rather remarkable, if long-winded and not entirely satisfactory attempt at deriving basic features of quantum theory from a minimal set of operational assumptions[19]. Destouches started from the idea that distinct measurements afford different resolutions of a given quantity $A$, associated with different partitions of the 'spectrum' of its admissible values. It is assumed preparations exist, for which measuring $A$ with a given resolution yields an outcome that falls with certainty within a given interval $E_i^A$. Destouches assigns to any such preparation a predictor[20] $X_i^A$, such that the probability $p_{X_i^A}(E_i^A) = 1$, and a suitable set of $X_i^A$ can be chosen so as to form an orthogonal basis of an appropriate 'predictor space'. Owing to this structure, any preparation to which a measurement of $A$ at the given resolution is relevant can be assigned a predictor, in the form of a vector $X$ that can be expanded in the $\{X_i^A\}$ basis, with the requirement that $p_X(\cup_i E_i^A) = 1$. Destouches's next essential move consists in assuming that, for any quantity $A$, the algorithm for computing the probabilities of measurement outcomes given a '$X$ preparation' should be independent of the chosen partitioning of the spectrum of $A$, i.e. of the choice of basis[21]. The upshot is a probability function of the form $f(x) = |x|^k$, where $x$ is a coefficient in a suitable linear expansion of the predictor $X$. The expression of $f$ follows from a consistency requirement that implies the satisfaction of the Cauchy functional equation

$$f(|x||y|) = f(|x|)f(|y|).$$

Destouches leaves it as an empirical matter to determine the value of *k*. However, a formal proof that *k* must be equal to *2* was supplied a few years later by P. Destouches-Février[22]. The proof, which relies on a theorem by Birkhoff and von Neumann[23], boils down to ensuring consistency between the projective (inner product) structure of the predictor space and the requirement that probabilities add up to 1.

---

[19] Destouches 1942.
[20] Destouches uses the expression '*élément de prévision*'.
[21] Somewhat similarly, Saunders (2004) has recently shown that, if the rule for computing expectation values depends only on conditions imposed on a preparation, then requiring that those expectation values be the same regardless of differences in 'descriptions' of the corresponding experiments is sufficient for constraining the rule to be Born's.
[22] Février 1946, 1956.
[23] The projective character of a framework with a built-in nondistributive orthocomplemented lattice structure implies as well as requires the existence of an involutive anti-isomorphism over the reference field – the complex number field in the case of quantum theory (Birkhoff and von Neumann 1936, reprinted in Hooker 1975, p.14).

Several decades after Destouches, other significant attempts at a partial reconstruction of quantum theory were, in particular, made independently by Y. Tikochinsky[24] and A. Caticha[25]. The idea of both is to assign numbers (real or complex) as an auxiliary device in the process of working out the probability of 'transition' between an initial (*I*) and a target (*T*) configuration[26]. (*I*) refers to an experimentally given or theoretically assumed preparation and (*T*) to the end product of an observation or measurement (whereby a new preparation effectively obtains). As sequences of transitions, mutually exclusive 'parallel' transitions, or both, are considered, assigned numbers – amplitudes – must combine in a consistent fashion. The relevant consistency constraints amount to requiring associativity and distributivity, and they result in an essentially unique form for the composition rules: those are the sum and product rules ('Feynman rules') which, in standard quantum theory, respectively apply to amplitudes associated with parallel and successive transitions. Given the assumption that the corresponding probabilities are mutually independent, the product rule for amplitudes then leads[27], for the probability $p(I \to T)$ as a function of the total amplitude, to what is in effect the same Cauchy relation that is crucial to Destouches's derivation, hence $p(I \to T) = |x|^k$, where *k* is a positive real number. Identity of the endpoints in the 'self-connection' $I \to I$ warrants[28] setting $p(I \to I) = 1$ as the probabilistic expression of a tautology. Breaking $I \to I$ down into a complete set $\{K_i\}$ of alternative subtransitions, given the sum and product rules for amplitudes $a(X \to Y)$, leads to *k=2* through the matching of $p(I \to I) = |\sum_i a(I \to K_i)a(K_i \to I)|^k$ with the requirement that the probabilities of all mutually exclusive subtransitions add up to 1. Allowing amplitudes to be complex, the amplitude for the inverse transition that is obtained by exchanging endpoints is shown[29] to be equal either to the amplitude itself or to the complex conjugate of that amplitude. The former is ruled out because there would then be choices of amplitudes for which $\sum_i a(I \to K_i)^2$ vanishes (this is the nearest Tikochinsky comes to justifying the occurrence of *complex* amplitudes in quantum theory). The value *2* of *k* is, there again, found to reflect consistency between the additivity of probabilities and the projective structure of the predictor space[30].

Let the two 'kets' $|\Psi^{(1)}\rangle$ and $|\Psi^{(2)}\rangle$ be predictors that are assigned respectively to two preparations $\mathbf{P}^{(1)}$ and $\mathbf{P}^{(2)}$, and *A* be an observable, the measurement of which is used to distinguish (statistically) between the two preparations – think about one using the relative numbers of Heads and Tails outcomes in repeated tosses of two (not necessarily

---

[24] Tikochinsky 1988a,1988b.
[25] Caticha 1998.
[26] The word must be understood here without any spatial connotation.
[27] Tikochinsky 1988b.
[28] There is no need to introduce, as Tikochinsky (1988b) does, time-reversed 'transitions' in order to ensure that both endpoints refer 'to the same time': temporal considerations are no more required for working out the probability rule than they had been needed for deriving the composition rules for amplitudes.
[29] Tikochinsky 1988b.
[30] As S. Saunders points out about his own derivation of the Born rule, "examination of the proof shows that the dependence of probabilities on the modulus square of the expansion coefficients of the state ultimately derives from the fact that we are concerned with unitary evolutions on Hilbert space, specifically an *inner-product* space, and not some general normed linear topological space." (Saunders 2002)

fair) coins in order to evaluate the 'statistical distance' between them. Given preparation $\mathbf{P}^{(i)}$, the probability of the result $a_k$ associated with eigenvector $|a_k\rangle$ of $A$ is $p^{(i)}(a_k) = |\langle a_k|\Psi^{(i)}\rangle|^2$. In the context that is associated with choosing $A$, the expression of the statistical distance between the two preparations is then[31]

$$d_A(\mathbf{P}^{(1)}, \mathbf{P}^{(2)}) = cos^{-1}(\sum_{i=1}^{N} |\langle a_i|\Psi^{(1)}\rangle||\langle a_i|\Psi^{(2)}\rangle|)$$

Two preparations that correspond to the same eigenket of $A$ are thus trivially indistinguishable, whereas they are statistically as far apart as two preparations can be if the corresponding predictors are two distinct eigenvectors of $A$. The sum reduces to a single term if one of the preparations corresponds to an $A$ eigenket assignment. The statistical distance is then identical with the Hilbert space angle between the rays associated with those preparations. This requires a suitable reference observable to be chosen, relative to which a given preparation is statistically gauged. Insofar as it requires choosing a reference set of mutually compatible observables (e.g. all of those that commute with $A$), this connection between statistical and angular distance can be regarded as an expression of context-dependence. As for the relative phases of amplitudes, their justification must be looked for in transformational, group-theoretically regulated properties that determine the form of the operator representatives of relevant, e.g. spin observables and their mutual relationships[32].

The 'statistical algorithm of quantum theory[33]' (SAQM) can also be formulated without any mention of amplitudes, e.g. in terms of probability tables[34] and rules for extracting from any such table the probability of a given outcome. This requires choosing a set of reference measurements in terms of which the probability tables can be set up. Such measurements are 'mutually unbiased' just in case, if experimental conditions are such that measuring one observable in the set is certain to yield a given outcome, then all the possible outcomes of measuring any other observable in that set are equally likely. An example of such a set is given by the spin-½ observables $\sigma_x$, $\sigma_y$ and $\sigma_z$ that correspond to mutually orthogonal directions $x$, $y$ and $z$. Observables in such a set are as statistically distant as comparable types of measurement can be, so that redundancy in the information supplied by measuring any two such observables is minimal. The observables in question can then be regarded as providing an optimal reference for setting up a probability table and deriving from it a density matrix. If $N+1$ mutually unbiased basis sets are available, a $N \times N$ density matrix can be constructed from the probability table and, conversely, the probability table is uniquely derived from the density matrix. The table is only constrained by the requirement that the probabilities of the outcomes of measuring any given reference observable add up to *1*, and that the density matrix derived from that table admit no negative eigenvalue. The same conditions have to be satisfied in the composite case. If it is additionally required that the probabilities of the outcomes of reference measurements performed on subsystems should not depend on measurements one may (choose to) perform on the other 'complementary' subsystems within the composite, then it follows that $N^2 - 1$ real numbers are needed for setting up a complete probability table, and this number is just the same as that needed to specify a density matrix in the 'full' tensor product space. The physical (ontological) neutrality of the basic requirements is clear enough, as is the fact that, claims of nonlocality notwithstanding, all of the predictions of quantum theory in the composite case can be derived from such tables.

---

[31] Wootters 1981.
[32] See Lévy-Leblond 1974 for an enlightening discussion.
[33] Redhead 1987, p.5.
[34] Wootters 1986.

An important aspect of the above derivations of both the rules for amplitudes and that for probability is that they are essentially independent of assumptions regarding the nature – 'quantal' or whatever – of the systems to which the predictions apply. That much is also true of Lucien Hardy's tightly worked out derivation[35] of the mathematical backbone of quantum mechanics from four or five basic axioms. Hardy's *ab initio* reconstruction leads him to conclude that "quantum theory, when stripped of all its incidental structure, is simply a new type of probability theory[36]." In fact, switching from the SAQM to a vector space realization of classical probability theory comes down to accepting or rejecting a mere continuity requirement.

Four only of Hardy's axioms actually contribute to the derivation. The remaining one (*HA0*) serves only the purpose of specifying how computed probabilities relate to the collection of relevant data: this is achieved, in practice, through the identification of the probabilities of measurement results with limiting relative frequencies. Nothing, however, in Hardy's subsequent derivation of the SAQM depends on his preference for a frequency-based understanding of probability[37]. Hardy's approach builds upon the idea that a given setting of an appropriate preparation device fully determines the probabilities of all 'non-null' outcomes of measurements one may perform on its output. It is reasonable to assume that there exists a minimum number[38] $K$ of probabilities, knowledge of which suffices to characterize the predictive yield or 'potential' of a given preparation. This yield is encapsulated in what Hardy chooses to call a *state*, presumably to make the connection to quantum theory as explicit as possible from the outset. This choice of terminology is misleading, however, for it will soon clearly appear that the framework Hardy works out is primarily aimed at yielding probabilistic predictions, whatever the subject matter. This should be borne in mind, given Hardy's common reference to "systems in some state[39]" or his speaking of "ascrib[ing] a state to a preparation[40]".

A 'Hardy state' is entirely specified through the listing of $K$ outcome probabilities, subject to completeness and closure conditions. Hardy now takes the crucial step of writing those probabilities as components of a $K$-dimensional vector $\boldsymbol{p}$. 'Pure states[41]' are by definition those primary vectors that cannot be written as convex sums of other vectors in the relevant set. Hardy regards it as a "driving intuition" that such "pure states represent definite (non-probabilistic) *states of the system*[42]". However, such an ontological construal of 'states' detracts from the operational foundation of the formalism, where the whole point of introducing them is to provide a convenient linear representation of probability distributions for measurement outcomes (there is no question here of hazily conceived propensities of physical or other kinds of systems). In effect, what those 'pure states' provide are reference sets in the following sense: to any such pure state, there corresponds a vector $\boldsymbol{p}_j$, the $j^{th}$ component of which is set equal to *1* whilst all others are *0*. This is associated with a setting of the preparation device for which the $j^{th}$ outcome is certain to obtain if the relevant test is actually performed. Alternatively a vector $\boldsymbol{r}_j$, such that $\boldsymbol{r}_j.\boldsymbol{p}_k=\delta_{jk}$, can also be assigned to the corresponding 'yes-no measurement'. Generalizing to an arbitrary preparation and any choice of measurement, the outcome probability is given by the scalar product $\boldsymbol{r}.\boldsymbol{p}$ of suitably chosen $\boldsymbol{r}$ and $\boldsymbol{p}$ vectors. That Hardy's states can be indifferently[43] represented as $\boldsymbol{p}$ or $\boldsymbol{r}$ vectors bears further witness to their being close relatives to Destouches's predictors and correspondingly devoid of ontological significance.

---

[35] Hardy 2001a,b.
[36] Hardy 2001b, p.1.
[37] Hardy's introduction of 'mixed states' as linear combinations, with suitable *probabilistic* 'weights', of what he calls pure states (see text below), suggests that the primary view of probability that (*HA0*) expresses is overlaid with 'subjective' aspects. However, a Bayesian alternative to (*HA0*) (see Schack 2002) is unlikely to satisfy many physicists, reluctant as they are to accept in their field an epistemic conception of probability (i.e. of probability as 'degree of belief'); either because they feel it is simply inadequate in the context of handling empirical data, or for fear that their ideal of objectivity might be threatened.
[38] Hardy's reference to $K$ as a "number of degrees of freedom" suggests a parallel with classical mechanics that is unwarranted – indeed almost incongruous – in this context, and is therefore best avoided.
[39] Hardy 2001b, p.2.
[40] *Ibid.*

Assuming that there is a set of 'states' whose members can be distinguished from each other by a 'single shot' measurement, there is no reason a priori for expecting their minimum number $N$ to be equal to $K$. Nonetheless, a structural relation presumably exists between those two numbers; a relation Hardy ascribes to "a certain constancy in nature[44]". Regardless of whether nature has any relevance to the matter, it is a sensible prerequisite for setting up an optimal framework for prediction that the predictive yield of a preparation, as encapsulated in a 'predictive vector', should properly (functionally) connect to distinguishability within an outcome set. The (*HA1*) axiom asserts this connection, with the additional ('simplicity') requirement that $K$, as a function of $N$, take the smallest value that is consistent with the full axiom set.

If the mode of preparation, or any conditions that prompt the assignment of a given predictor, imply that certain measurement outcomes are precluded, the maximum number of distinguishable states, hence the dimension of the predictor space, can be reduced accordingly. This axiom (*HA2*) conforms to the expectation that the probability of occurrence of an outcome should not depend on whether the set it belongs to is embedded in some larger set. Axiom (*HA3*) then addresses those cases where "a preparation device ejects its system in such a way that it [the system] can be regarded as made up of two subsystems[45]" (1) and (2). If $N_1$ and $N_2$ are the numbers of distinguishable Hardy 'states' that relate to the performance of measurements on (1) and (2) respectively, then the number of distinguishable 'states' associated with joint (1)-(2) preparations cannot be less than $N_1N_2$. (*HA3*) requires $N$ to be, in all circumstances, *equal* to $N_1N_2$. The $K_i = K(N_i)$ components of a vector $\boldsymbol{p}^{(i)}$ are measurement probabilities relative to the (*i*)-member (*i = 1* or *2*) of any (1)-(2) pair. The construction of a 'joint' probability matrix in which the outcome probabilities for measurements performed on (1) and (2) separately are compounded suggests that the number *K* of real parameters that are necessary and sufficient for defining a (1)-(2) 'state' is also just equal to the product $K_1K_2$ (this suggestion is also part of (*HA3*)): "there should not be more entanglement than necessary[46]" for prediction to be successfully (optimally) achieved.

---

[41] The null state is excluded from the set of pure states.
[42] *Ibid.* (italics added).
[43] *Ibid.*, 2.
[44] Hardy 2001a, 6.14.
[45] *Ibid.* 6.16.

(*HA2*) and (*HA3*) entail that $K(N+1) > K(N)$ and $K(N^2) = K^2(N)$, and the only polynomial in *N* that satisfies those conditions is $K(N) = N^r$. Now, in quantum theory, if the relevant Hilbert space is of dimension *N*, then $N^2 - 1$ independent real numbers are required to specify a density matrix (or, in usual parlance, to fully determine a 'state'). This is also the number of independent entries in Wootters's probability table, which was seen to follow from requiring, besides the normalization of probability, that the probabilities of the results of measurements performed on any subsystem should not depend on which measurement is actually (or intended to be) performed on the complementary subsystem(s).

Insofar as it does comply with Wootters's 'local accessibility' thesis ("any set of measurements which are just sufficient for determining the states of the subsystems are, when performed jointly, also just sufficient for determining the state of the combined system[47]") quantum theory *qua* predictive framework optimally uses information supplied by measurements performed separately on subsystems. It is satisfying that Wootters's conjecture: that any theory which satisfies local accessibility through the optimality condition[48] $g(N_1N_2) = g(N_1) + g(N_2) + g(N_1)g(N_2)$ would be such that $g(N) = N^r - 1$, with *r* a positive integer, happens to be vindicated by Hardy's (*HA1-2-3*) axioms, since those imply that $K(N) = g(N) + 1 = N^r$. Given the simplicity requirement that (*HA1*) includes, we are left with only two kinds of scheme, depending on whether *r* equals *1* or *2*. The first one (*K = N*, *r = 1*) corresponds to a vector space realization of the classical probability calculus, in which the number of parameters needed to probabilistically characterize a preparation is equal to the maximum number of distinguishable 'states'. Suppose, for example, that the set-up includes an apparatus that releases either red or green balls. The *K = 2* probabilities of finding a ball to be red or to be green exhaust the specification of the 'state', whilst the maximum number of distinguishable 'pure states' is obviously *N = 2*: a single red/green measurement will at most distinguish between those two colors. It is, however, conceivable that the equality of *K* and *N* might not always be satisfied, and the SAQM, for which $K = N^2$, appears to be a case in point. *Real* Hilbert spaces are actually ruled out[49] by (*HA3*) : for those spaces $K = \frac{N(N+1)}{2}$ and specifying 'states' in the bipartite case would then require more ($K > K_1K_2$) than can provide data gathered separately on the two subsystems[50], thereby violating local accessibility. Complex numbers, therefore, appear to be a necessary ingredient in the kind of linear predictive scheme Destouches tentatively initiated, of which Wootters hypothesized some essential features and which Hardy successfully derives from his simple set of axioms.

---

[46] Hardy 2001b, p.10.
[47] Wootters 1990, p. 44.
[48] Wootters 1990.
[49] For a counterexample, see Wootters 1990, p.44.
[50] Hilbert spaces over the quaternions – if one should ever care to consider such possibilities – would require strictly less ($K < K_1K_2$).

More light has recently been shed on the relationship between complex numbers and $\otimes$-composition. A linear operator $\sigma$ has been shown[51] to exist, such that the joint probability for any ordered pair ($E,F$) of locally 'observable' positive operator-valued measures takes the trace form $tr(\sigma(E \otimes F))$. $\sigma$ turns out to be unique in the case of complex Hilbert spaces (the proof falls short of establishing that $\sigma$ is, or has to be, a density – statistical – operator). Uniqueness follows because the set $\{E_\mu \otimes F_\nu\}$ forms a complete basis for Hermitian operators. It is as yet unknown whether this only holds in the complex case, although it is highly likely: when the field is complex, the operator space of the tensor product is known to be isomorphic to the tensor product of the original operator spaces. By contrast, the dimensionality of the space of symmetric operators on a real Hilbert space is strictly less than that of the (complex) vector space of Hermitian operators over the base space, thus preventing $\sigma$ to be uniquely specified.

If normalization is not assumed, all of the predictively useful information a $N$-dimensional density matrix $\rho$ encapsulates amounts to that which is supplied by a Hardy vector whose components are $K = N^2$ probabilities. The equivalence holds because any Hermitian operator which admits a $N \times N$ matrix representative can also be written as a linear combination, with real coefficients, of $K = N^2$ projection operators. Writing those $K$ operators as the components of a vector $E$, then $p = tr(\rho E)$. The most general expression of probability is $tr(\rho \mathbf{A})$, where $\mathbf{A}$ is a positive operator such that[52] $\mathbf{A} = r.E$.

Whatever essential difference there is between 'quantal' and 'classical' frameworks boils down to the composition of the sets of allowed $p$ and $r$ vectors. The split occurs with axiom (*HA4*), which asserts the existence of a continuous reversible transformation along a path connecting two arbitrary 'pure states'. This is directly linked to 'superpositions' of states being allowed in the $r = 2$ case: a Hardy state can then be gradually (continuously, reversibly) transformed into another, and the 'in-between' states are linear combinations of those two states (the continuous distribution of points on the surface of the Bloch sphere illustrates that property of quantum-theoretic 'states' in the two-level case). In contrast, 'classical' ($K = N$) pure Hardy states form a discrete set (compare the red and green ball example). It is somewhat ironical that *classical* probability theory should be characterized by a necessity to 'jump' between 'pure states', whereas the trait that singles out quantum theory among the schemes that satisfy Hardy's other axioms would be the existence of continuous transformations between its 'states'. Hardy suggests that the necessity to 'jump' from a 'classical state' corresponding to, say, a ball *being* in one box to that corresponding to its *being* in another merely reflects our crude partitioning of possibilities into what just *appears* to us to be clear-cut alternatives. It is quite hard, to say the least, to figure out how certain human (perceptual?) limitations could determine the type of Hardy scheme that should be selected as adequate. Why should a $K = N^2$ framework replace its 'classical' $K = N$ counterpart as our means of investigation 'reach out' to a (sub)atomic realm 'where' our perception cannot guide us and our rational expectations are thus likely to be challenged? This is, at any rate, idle speculation: nothing in the content of Hardy's axioms warrants ontological preconceptions about systems and their 'modes of being', let alone invocations of 'natural' limits to our cognitive or perceptual capacities.

---

[51] Fuchs 2002.
[52] Hardy 2001a, 5 & 8.7.

The most general kind of time evolution that is consistent with all the axioms is a linear and completely positive map on the space of operators on Hilbert space, such that the trace (or normalization coefficient in the pure case) does not increase. Unitary evolution obtains if it is required that the corresponding transformation should also be invertible and that it should preserve the trace, whereas 'reduction' means net trace decrease, which is accompanied by the increase of the von Neumann entropy. Updating the representation upon acquisition of new information is subject to the same basic constraints in classical (probability) and quantum theory. However, the updating rules reflect the structural features of each particular framework[53]. Hardy's axioms – invocation of 'states' notwithstanding – provide no grounds for believing that any kind of physical 'collapse' accompanies the updating process.

*3.    Prediction and beyond*

> "…I can safely say that nobody understands quantum mechanics… Nobody knows how it can be like that."
> – *R.P. Feynman*
>
> Prédire n'est pas expliquer.
> – *R. Thom*

None of the derivations of the SAQM we have reviewed hinges on any assumption regarding the nature, 'quantal' or 'classical', of physical systems. A probabilistic formalism of the same kind might well turn out to be applicable, with benefit, to disciplines that have little, if anything, to do with the concerns of physicists[54]. One might also maintain that operational effectiveness in anticipating results of measurements is all that one should ask of a physical theory, and therefore all that quantum mechanics can realistically provide. Asking for 'more' would not be sensible, confusing as it does the anticipative and predictive aims of science with 'metaphysical' yearnings that cannot be substantiated. Many, however, will certainly side with Einstein, who once expressed his opinion on the matter in a pretty harsh way: "If that were so then physics could only claim the interest of shopkeepers and engineers; the whole thing would be a wretched bungle[55]." Physics, indeed, would not be half as exciting if all it could offer were a laundry list of 'results' and a compendium of useful recipes. But however strongly one may concur, this is hardly an argument; merely the negative of a widespread *aspiration*: that of gaining an ever more faithful and accurate understanding of 'what there is'.

---

[53] Despite appearances, 'state vector collapse' may not radically differ from the ordinary Bayesian update of 'classical' probabilities; see Fuchs 2002.
[54] See Aerts and Gabora 2004 for an application of the same kind of Hilbert space-based scheme to the classification and combination of concepts.
[55] In a letter to Schrödinger, quoted from Przibram 1967, p.39.

Let's try to overcome our reluctance to surrender those uplifting, if unrealistic prospects and cautiously assume that a physical theory cannot simply (we do not say 'simply cannot') provide a description, be it approximate and forever amendable, of 'things as they are' or of 'the world as it is'– no more, indeed, than our brains provide any *direct* access to a reality out there. One might argue that such a theory may nonetheless adequately reflect fundamental aspects of *our experience*, bearing in mind that the latter is never 'raw' but constantly and necessarily 'informed' through those feed-backs of our mental make-up that are necessary for converting 'unformatted' sensory inputs into reliable and shareable expressions of our thought. According to M. Bitbol[56], "the basic formalism of quantum mechanics can effortlessly be construed as a structural presupposition of any activity of production and unified anticipation of mutually incompatible contextual phenomena[57]." A correct 'philosophical' evaluation of quantum mechanics, Bitbol contends, should have the quasi-therapeutic effect of dispelling a major illusion: that a phenomenon can always be in principle detached from the very conditions that make its occurrence possible. What Bohr strove to express in his later writings[58] is that phenomena like those one encounters in atomic physics are necessarily and indissolubly co-determined by the experimental conditions of their manifestation ('context-dependence'). This belated recognition of the aim and structure of quantum theory would

> "undermine the pictures so cherished by supporters of the ontological (disengaged) outlook…by showing that the predictive success of some of our most general scientific theories can be ascribed, to a large extent, to the circumstance that they formalize the minimal requirements of any prediction of the outcomes of our activity, be it gestural or experimental. The very structure of these theories is seen to embody the performative structure of the experimental undertaking[59]."

Once this is realized, there will "no [longer be any] need to further explain" the adequacy of the quantum rules "by their ability to reflect in their structure the backbone of nature[60]". The position Bitbol defends is a contemporary offshoot of a distinguished 'critical' tradition[61], the advocates of which have striven to provide a middle way between the unreasonable expectations of unrestrained realism on the one hand, and the more unpalatable aspects of instrumentalism on the other.

---

[56] Bitbol 1996,1998,2002.
[57] Bitbol 2002.
[58] For a lucid discussion of the evolution of Bohr's views, see Held 1994.
[59] Bitbol 1998.
[60] *Ibid.* Physicists or philosophers with realist leanings may be "very imprudent", according to Bitbol, in believing that "self-existent objects are what justify the intentional attitudes". An unprejudiced philosophical inquiry would rather prompt the conclusion that "[the] project of ontologizing certain theoretical entities appears a mere attempt at hypostatizing the major invariants of those activities." [Bitbol 1998]
[61] This line of thought originates in Kant's *Critique of Pure reason*. Modern updates, like that of Hintikka (1999) to which Bitbol is indebted, downplay Kant's much-criticized apriorism and redirect focus on 'structural' constraints of experimental activity.

Bitbol bases his pronouncements almost entirely upon (i) Destouches's derivation of the Born rule (see Section 2) and (ii) Heelan's unconventional take on so-called quantum logic[62]. The sole purpose of $\psi$ functions, kets and so forth would be to provide sets of abstract predictors, whose mathematical properties would basically reflect their invariance under changes of 'experimental contexts'. If the mutual 'compatibility' of experimentally ascertainable quantities is restricted to non-equivalent classes, propositions that can be asserted about values of the quantities in question cannot be co-ordinated in a unified way within a Boolean framework, isomorphic to the algebra of set-theoretical operations. Instead, the validity of Boolean logic can be maintained only within strictly delimited contexts. Predictive consistency across such non-equivalent contexts will then call for a suitable 'metacontext' logic. Heelan[63] tentatively argues that

> "the locus of nonclassical logic in quantum mechanics is in the plane in which physical contexts are related to one another, and not, as all writers have hitherto held, in the plane of single quantum-mechanical events…the proper subject matter of so-called quantum logic would be the manifold of experimental contexts in which it is relevant to use one linguistic or conceptual framework rather than another[64]".

The characteristic orthocomplemented and non-distributive lattice structure of quantum logic would consist in a partial ordering of a set of experimental languages that are pairwise incompatible. Each language would correspond to the selection of a definite context i.e. of a class of experimental propositions whose conjunction is operationally meaningful. A *meta*context language would then be required in order to co-ordinate propositions of languages that are tied to distinct contexts, and the corresponding lattice structure would be that of quantum logic.

*Ab initio* derivations of the SAQM (cf. Section 2) suggest that quantum theory owes its structure, and in particular the form of its most basic rules, to its capacity to operate, as a predictive scheme, both within and across such non-equivalent, 'incompatible' contexts. Acknowledging this connection between the role of the SAQM and its structure would "automatically defuse[s] major paradoxes[65]."

Is this all that one should expect from a fundamental physical theory? Bitbol's answer is unambiguous: "the only thing a physical theory does, and the only thing it has to do, is to embed documented actualities in a (deterministic or statistical) framework, and to use this framework to anticipate, to a certain extent, what will occur under well-defined experimental circumstances[66]." Our most successful theoretical frameworks would be (nothing but) "embodiments of the necessary pre-conditions of a wide class of activities of seeking and predicting[67]." Nevertheless, rather than seize an opportunity to reassess pre-conditions of our activities within a world-as-experience(d), many are expected to remain stuck with their classical, 'pre-critical' illusions and "systematically favor a disengaged outlook, even though their very undertaking is grounded on the presuppositions of an engaged activity[68]."

---

[62] Heelan 1970a,b.
[63] Heelan 1970a,b.
[64] *Ibid.*
[65] *Ibid.*
[66] Bitbol 1998.
[67] *Ibid.*

But the question remains of why it should have taken confrontation with the puzzles of atomic physics for physicists to realize – to the extent they did – that means and procedures of investigation cannot be 'neutralized' in all circumstances. It is rather hard to believe that the pioneers of quantum physics did, through the finiteness of Planck's constant and subsequent developments, unwittingly stumble upon (nothing but) the empirical trace (?) of – to put it in Bitbol's philosophese – pre-conditions to the co-ordination and anticipatory effectiveness of experimental activity, ending up with a 'mechanics' that embodies in its structure a major philosophical lesson. Besides, how terms get reinforced or cancel out in the process of computing probabilities with the SAQM is certainly 'wavelike' in a rather metaphorical sense: this is paper interference, part and parcel of the operation of a particular type of formalism that allows one to predict the measured values of measurement outcomes, cross-sections and so on. It is quite another matter, however, to claim that a basic mathematical trait associated with consistent operation of a predictive scheme 'across' different contexts should be underwritten by specific interference patterns exhibited e.g. on photographic plates.

It might well be that the SAQM is all we can afford, whether in principle or 'only' in practice, as a framework in which our – necessarily mediated – approaches to the microworld can find a consistent and effective mathematical expression. This should not prevent us, however, from attempting to trace the quantum back to sound and compelling physical principles. Granting that the predictive rules of quantum theory can, as we saw, be justified without making any 'ontic' reference to physical systems, I shall put forth the following suggestion: far from falling from the sky, the SAQM actually arises as a result of selecting a certain type of mathematical structure. Historically, this 'selection' unwittingly occurred in the process of elaborating a theoretical framework for atomic physics. More precisely, quantum theory will, in the next section, be seen to emerge as a by-product of setting up a consistent 'protoquantal' dynamical framework, based on the acceptance of a definitely 'non-classical' but well-motivated (and by no means preposterous) postulate. Given the assumption that quantum mechanics, in its currently accepted form, turns out to be just such a by-product, reasons will be adduced for resorting to the SAQM, beyond the historical fact that the protoquantal framework in question was never suspected by the great pioneers of quantum physics.

---

[68] *Ibid.*

*4. From the Quantum to the SAQM*

In the early days of quantum physics (mid-1910s), some parallels were pointed out between the integral formulation of the Bohr-Sommerfeld quantization condition and the action-angle methods that had been introduced half a century before to help tackle problems in celestial mechanics[69]. Such similarities, just as the dimensions and apparent ubiquity of Planck's constant itself, suggested focusing upon *action* as the key quantity for unlocking the mystery of the quantum and working out a compelling alternative to classical mechanics. Such hints were all but forgotten, however, once Heisenberg, Born and Jordan had come up with an effective 'matrix' framework that seemed to owe nothing, except for some dubious analogies, to any classical predecessor. Action and related issues had briefly come back to the fore with the contributions of Louis de Broglie and Schrödinger. Nevertheless, the view eventually prevailed that no amount of reflection on the basic concepts and assumptions of classical mechanics could possibly help understand how and why quantum physics is so 'special' in the ways it departs from 'classicality'.

It is well known that, if one substitutes into Schrödinger's time-dependent equation a polar form of its solution $\psi$, one of the resulting two equations turns out to be identical with the classical Hamilton-Jacobi equation, except for the presence of an 'extra' contribution. The de Broglie-Bohm 'pilot-wave' approach[70] focuses on that 'non-classical' term, interpreted as a new kind of potential, which is regarded as the source of all 'quantum effects'. However, it is a major drawback of that approach that it relies on the prior availability of the Schrödinger equation. Besides, although Bohm's guidance condition suggests a trivializing map whereby phase space coordinates transform into a set that is constant in time, the dual of that transformation appears to be left out without reason[71]. It is as if the initial substitution pointed toward some possible alternative to Hamilton-Jacobi theory, but without giving more, as it were, than half a hint. The dependence of Bohm's quantum potential on Hamilton's principal function also implies higher derivatives of that function. What is the structural role and conceptual significance of such higher derivatives?

In the first part of his 'wave mechanics' paper[72], it is with the purpose of addressing quantization as an eigenvalue problem that Schrödinger expressed the 'characteristic function' as the logarithm of a new function $\psi$. Conversely, writing the $\psi$ function, solution to the (one-dimensional) stationary Schrödinger equation, in polar form leads to a third-order non-linear differential equation that must be satisfied by the quantum-mechanical version of the characteristic function $S$:

$$\frac{1}{2m}(S')^2 + (V - E) = -\frac{\hbar^2}{4m}\left[\left(\frac{S'''}{S'}\right) - \frac{3}{2}\left(\frac{S''}{S'}\right)^2\right]$$

---

[69] See Jammer 1989, pp 103-107.
[70] Bohm 1952, Holland 1993.
[71] Holland 2001a.
[72] Schrödinger 1926.

($S' = \frac{\partial S}{\partial q}$ etc.). This equation differs from the stationary Hamilton-Jacobi equation by its non-vanishing right-hand side. The function $S$ can be interpreted as the generator of the motion for a single trajectory ($S'$ then corresponds to the conjugate momentum). Knowing the trajectory then suffices to specify the $\psi$ function – there is no need for an ensemble of such trajectories to be considered. The $\psi$ function, however, does not always afford resolution into single trajectories. Most significantly – this can hardly be regarded as a 'coincidence' – the 'extra' term turns out to be proportional to the so-called Schwarzian derivative $\left(\frac{S'''}{S'}\right) - \frac{3}{2}\left(\frac{S''}{S'}\right)^2$ of $S$.

The above equation will, from now on, be referred to as the *quantum stationary Hamilton-Jacobi equation* (QSHJE). The QSHJE has never been fashionable as a conceptual lead to the resolution of quantum mysteries. Worth mentioning, however, is Floyd's 'trajectory representation of quantum mechanics[73]', which proceeds directly from the QSHJE, bypassing the Schrödinger equation. Floyd's approach differs from Bohm's in some fundamental respects, not least in the nature of the trajectories. Conjugate momentum is not identified with mechanical momentum and, since the QSHJE itself completely determines the motion, there is no need for any guiding field. Floydian trajectories, though consistent with the requirement that probability should be conserved, are not distributed in accordance with the $\psi$ function density. They are also distinct from Feynman paths: in Floyd's account, the generator of the motion ought to be regarded as an alternative to Hamilton's characteristic function, whereas Feynman's propagator is classical. Nowhere is any randomness involved, but our general inability to resolve a $\psi$-based account into distinct trajectories happens (more about that later) to be 'fortunately' compensated by the use of the probabilistic methods of standard quantum theory.

Being able to derive the QSHJE *from* the Schrödinger equation is by itself of little help to figure out what set of theoretical requirements could give rise to *that* equation (QSHJE) in the first place, or to decide whether its solutions are more physically significant than 'wave functions'. Some invaluable light, however, has recently been shed on the matter by A. Faraggi and M. Matone[74], who succeeded in deriving the QSHJE on the basis of a single, theoretically well-motivated 'equivalence' postulate[75].

In a mechanical framework where the functional dependence of the characteristic function or 'reduced action' $S_0$ on a set of generalized coordinates $q$ determines the dynamics, does some coordinate transformation $q \to \hat{q}$ generally exist, such that the resulting reduced action $\hat{S}_0$ has the same dependence on $\hat{q}$ as $S_0$ has on $q$ ? If both $S_0$ and $\hat{S}_0$ are non-constant, such a coordinate transformation makes sense from the point of view of an observer in motion relative to both systems. However, no coordinate transformation can map a non-constant $S_0$ into the constant reduced action that characterizes, in classical mechanics, a free point mass with zero energy. Setting up an alternative ('non-classical') dynamical framework in which the existence of locally invertible coordinate transformations is not subject to any such limitation is precisely the aim of Faraggi and Matone. The programme is reminiscent of Hamilton-Jacobi theory, although coordinate transformations only are considered, the transformation of $p$ being induced by that of $S_0$ through $p = \partial_q S_0$.

---

[73] Floyd 1996, 2000.
[74] Faraggi and Matone 2000.
[75] For the sake of easy reference, the notation and terminology used in the following are those of Faraggi and Matone 2000.

Classical mechanics never treats canonical variables on a truly equal footing. However, the involutive nature of the Legendre transformation, which is implied by the definition of the generating function, gives rise to explicit $p - q$ duality (Hamilton's equations of motion are, up to sign, $p - q$ symmetrical). This duality can be made manifest by introducing a generating function $T_0$, which stands in the same relation to $q$ as $S_0$ does to $p$, i.e. $q = \partial_q T_0$. As a result, the description of a system can be indifferently achieved in either of the $S_0$ or $T_0$ pictures. Requiring both stability under evolution in time of the *S-T* relationship, and that symmetry be the highest amongst all possible interchanges of the two pictures leads to the 'natural' introduction of imaginary numbers (this is as far as the occurrence of complex numbers in quantum mechanics, at the dynamical level, can be traced back). Moreover, $S_0$ and $T_0$ are related through a so-called Möbius transformation (generators of the Möbius group are translations, dilations and inversions).

According to the *Equivalence Postulate* (EP), for any pair of dynamical systems *A* and *B* there exists a coordinate transformation $q^A \to q^B$ such that $W^A(q^A)$ transforms into $W^B(q^B)$, where $W = V(q) - E$ (*V* and *E* are the potential and the total energy terms). The postulate implies the existence of a coordinate mapping between any system and that corresponding to $W^0 \equiv 0$. If $S_0$ satisfies the classical stationary Hamilton-Jacobi equation (CSHJE), then the *W(q)*'s transform as quadratic differentials:

$$W'(q') = \left(\partial_{q'}q\right)^2 W(q).$$

Since $W^0$ can then only transform into itself, the *W* of an arbitrary system cannot be connected to $W^0$ by an invertible coordinate transformation. Therefore, implementing the Equivalence Postulate calls for a radical departure from classicality.

Mapping $W^0$ into some *W≠0* requires *W* to transform non-quadratically:

$W'(q') = \left(\partial_{q'}q\right)^2 W(q) + (q;q')$, where the inhomogeneous term $(q;q')$ accounts for the abovementioned departure. Besides, one should reasonably expect the succession of two transformations, one from $W^A$ into $W^B$ and another from $W^B$ into $W^C$, to be equivalent to a single coordinate transformation from $W^A$ into $W^C$. This 'pseudogroup' requirement amounts to the satisfaction of the cocycle condition

$$(q^A;q^C) = \left(\partial_{q^C}q^B\right)^2 [(q^A;q^B) - (q^C;q^B)],$$

which condition implies the invariance of $(q^A;q^B)$ under a Möbius coordinate transformation $q \to q' = \frac{Aq+B}{Cq+D}$. The $(q^A;q^B)$ term is then proportional to the Schwarzian derivative $\{q^A, q^B\}$, which is indicative of Möbius symmetry (the Schwarzian derivative $\{f,x\} = \left(\frac{f'''}{f'}\right) - \frac{3}{2}\left(\frac{f''}{f'}\right)^2$ is such that $\left\{\frac{Af+B}{Cf+D},x\right\} = \{f,x\}$). The cocycle condition uniquely determines the Schwarzian derivative, up to a global constant and a coboundary term, and $(q^A;q^B) = -\frac{\xi^2}{4m}\{q^A;q^B\}$.

The EP-based alternative to the CSHJE can, with no loss of generality, be written

$$\frac{1}{2m}\left(\frac{\partial S_0}{\partial q}\right)^2 + W(q) + Q(q) = 0$$

where $Q = -\frac{\xi^2}{4m}\{S_0, q\}$. The inhomogeneous term is in practice negligible for very small values of $\xi$, and the CSHJE is recovered as $\xi \to 0$ with $S_0$ the classical reduced action. Just as in classical mechanics, the dynamics is encoded in the functional dependence of $S_0$ on its argument ($S_0$ will still be referred to as a 'reduced action', bearing in mind that it is not identical with its classical namesake).

The resulting alternative to the CSHJE

$$\frac{1}{2m}\left(\frac{\partial S_0}{\partial q}\right)^2 + V(q) - E = -\frac{\xi^2}{4m}\{S_0, q\}$$

is just the QSHJE provided $\xi$ is identified with Planck's 'reduced' constant $\hbar$. Momentum is evaluated through $p = \partial_q S_0$ ($\neq m\dot{q}$), whereby a definite (Floydian) trajectory follows. The time-dependent Hamilton-Jacobi equation obtains upon substituting $S(q,t) = S_0(q) - Et$.

$S_0$ being a solution of the QSHJE is equivalent to $\psi$ and $\tilde{\psi}$, such that $e^{\frac{2i}{\xi}S_0} = \frac{\tilde{\psi}}{\psi}$, being linearly independent solutions of the equation

$$\left(-\frac{\xi^2}{2m}\frac{\partial^2}{\partial q^2} + V(q)\right)\psi = E\psi$$

Identification of $\xi$ with $\hbar$ yields Schrödinger's stationary equation.

The 'quantum potential' $Q$ is fundamentally different from Bohm's in that, consistently with the Equivalence Postulate, $S_0$ is never constant. This fact is reflected in the *bi*polar form

$$\psi = \left(\frac{\partial S_0}{\partial q}\right)^{-\frac{1}{2}}\left(Ae^{\frac{i}{\hbar}S_0} + Be^{-\frac{i}{\hbar}S_0}\right)$$

of a general solution of Schrödinger's equation. As Einstein objected to Bohm[76], the classical motion does not always get recovered in the (improper) $\hbar \to 0$ limit. The objection does not apply in the present context since the quantum reduced action can never be constant.

A consistent exploitation of the Equivalence Postulate requires the coordinate transformations to be locally invertible. Continuity conditions for the existence of the Schwarzian derivative must also hold on the extended real line. Schrödinger's eigenvalue problem connects to a local invertibility condition thanks to a theorem[77] that makes energy quantization a manifestation of the existence of a self-homeomorphism of the extended real line (the existence of this self-homeomorphism is directly reflected in the Möbius invariance of the ratio $\frac{\tilde{\psi}}{\psi}$ ). Thus, the EP leads to quantization as a response to symmetries that underlie the form of the QSHJE. Most significantly, quantization does not require that all 'physical' solutions to the Schrödinger equation should be square summable, but only that such solutions *exist*.

---

[76] See Holland 1993, p.243.
[77] Faraggi and Matone 2000, section 18.

Summarizing, the QSHJE can be derived as a consequence of complying with a single postulate (EP): the term whereby the QSHJE differs from its classical predecessor originates in the postulated universal existence of a trivializing coordinate transformation. Schrödinger's stationary equation follows as a result of linearizing the QSHJE. The existence of the trivializing coordinate transformation requires the form of a general solution of the Schrödinger equation to be bipolar, in accordance with Floyd's approach and in disagreement with Bohm's. An existence condition for the Schwarzian derivative is that the total energy *E* admit a value that is an eigenvalue of the corresponding Schrödinger equation, or equivalently that the Schrödinger equation admit a solution that is a $L^2(\mathbb{R})$ function on the real line. Quantization does not require that all 'physically acceptable' $\psi$ functions should be square summable. However, interest in solutions having that property is justified, *a posteriori*, insofar as their existence provides a structural basis for setting up the kind of effective predictive scheme that was discussed in Section 2. In its operation as such a scheme, quantum theory has amply proven to be able to make up for the potential loss of descriptive power implied in bypassing the Q(S)HJE and relying instead exclusively on solutions of the Schrödinger equation.

Are 'quantum' probabilities dispensable? Implications of a theorem by M. Gromov provide at least a hint at an answer. The theorem itself[78], which relates to the existence of a certain symplectic invariant (symplectic width), falls way beyond the scope of this paper. One of its consequences, however, is an intriguing classical precursor of Heisenberg's 'uncertainty principle'. Any attempt to determine empirically the phase space coordinates $(q_1, \dots, q_N; p_1, \dots, p_N)$ of a system with *N* degrees of freedom is, in practice, limited by the resolutions $\Delta q_i$ and $\Delta p_i$ of measurements. Letting $\varepsilon_{min}$ be the minimal value amongst the areas $\varepsilon_i = \Delta q_i \Delta p_i$ of all the *N* error 'boxes', the theorem implies that determining a pair $(q_k; p_k)$ with a given accuracy requires that some of the remaining degrees of freedom be known with the same or better accuracy, but the error cannot be made less than $\varepsilon_{min}$. In classical mechanics though, nothing prevents setting in principle $\varepsilon_{min} = 0$. It may be conjectured, however, that an alternative dynamical framework based upon the QSHJE preserves that consequence of Gromov's theorem, supplementing it with a lower bound $\approx \hbar$ on the permissible values of $\Delta q_i \Delta p_i$, where $\hbar$ is the empirical value of the structural constant $\xi$. Were this highly plausible conjecture to be proven[79], one might be justified in regarding the emergence of quantum theory as a response, however unwitting, to the impossibility of taking, even in principle, full advantage of a dynamical account based upon the QSHJE (an equation that is notoriously tricky to solve). The possibility of developing a Hardy *r=2* type of predictive scheme (SAQM) happens to be, fortunately enough, afforded through the existence of a 'special' class of solutions to the Schrödinger equation, which existence would manifest itself in the quantization of energy.

---

[78] Gromov 1985,1987; see Hofer 1998 for a survey. Thanks to Ivar Ekeland for drawing my attention to the implications of Gromov's theorem and supplying some references.

[79] This might connect to the result that the EP implies phase space nonlocalization, which directly reflects the simple fact that a point cannot be diffeomorphic to a line (Faraggi and Matone 2000).

If the QSHJE is regarded as the basic equation underlying all things quantum-mechanical, then classical dynamics can, in practice (only), be regarded as a limiting case of 'quantum behavior'. But why should a theoretical framework, in which the appearance of an action quantum $\xi = \hbar$ and the quantization of energy can[80] be conceived as manifestations of an 'equivalence' postulate, be adequate to investigating structures and processes on (sub)atomic scales? To the best of our current understanding, the most that can be said is to remark that the coincidence of $\xi$ with the empirically given $\hbar$ is neither more nor less mysterious than the identification of the structural constant $c$ with the empirical value of the speed of light[81]. All there is to the 'microscopic' adequacy, both descriptive (QSHJE, to the extent it is practicable) and predictive (SAQM), of quantum mechanics – besides some essential group-theoretical input – might just be the smallness of Planck's constant in conventional units that are tailored to human scales.

Early breakthroughs, from the resolution of the black-body or photoelectric puzzles to the Bohr-Sommerfeld model of the atom, had given physicists very few hints at structural necessities behind the occurrence of Planck's quantum. When a new scheme eventually emerged, its agreement with the empirical data appeared to be quasi-miraculous. As we have come to realize, this was made possible because the very form of the Schrödinger equation (or its matrix/operator equivalents) happens to lend itself to setting up a consistent and adequate 'non-classical' framework for prediction. Behind the 'miracle', there is the fact that Hilbert space structure both (i) yields, through a linearization procedure, a faint echo of protoquantal dynamics as captured in the QSHJE, and (ii) constrains the linear representation of groups that determine[82] the form of quantities whose measurement are the object of prediction by means of the SAQM. The latter is a Hilbert space-based (Hardy $r=2$) scheme for the calculation of probabilities of measurement outcomes that hinges on the identification of a cross-contextual mathematical entity or predictor ('state vector', density matrix) that characterizes the 'predictive potential' of a given preparation. To be adequate, the predictive algorithm must obviously preserve the compliance of probability with the Kolmogorov axioms within any given context i.e. with respect to any chosen set of mutually compatible observables. On the other hand, its basic rules will be constrained in accordance with an underlying metacontext lattice structure that is non-Boolean (Heelan's thesis). Although neither Hardy's axioms nor Destouches's more rudimentary *calcul des prévisions* make explicit reference to context-dependence, it is satisfying to conceive of the form of the statistical algorithm of quantum theory as reflecting its role in co-ordinating probabilistic valuations at a metacontext level – in other words, to think of it as an algorithm that functions both *within* and *across* definite contexts.

---

[80] Admittedly, much remains to be clarified about the conceptual meaning and the significance of the EP. What matters to us here is *not* any particular commitment to Matone and Faraggi's EP approach, but the recognition of the action-based roots of quantum mechanics and of the central role of the QSHJE as the physical source of the Schrödinger equation.

[81] It will never be emphasized enough that the Lorentz transformation can be (and, I strongly believe, should be) derived from a single postulate, without making any reference to the propagation of light (see e.g. Lévy-Leblond 1976).

[82] Lévy-Leblond 1974.

Tensor product composition allows the predictive formalism to be applicable to preparations that are multipartite, i.e. such that various agents may operate separately on 'parts' of a 'larger' system, then work out statistics of their results, including specific correlations. It is legitimate to inquire about the way mutual relationships between parts of a composite system should be treated in a truly 'quantum dynamical' framework, and how the dynamics of the whole connects to that of the constituent parts. However, those are matters that will not be settled using the resources of quantum theory as we know and use it, if the latter essentially functions as a predictive scheme. Careful comparison of locally gathered data does exhibit correlations that cannot be accounted for on the basis of a simple-minded picture relying on common causes. Owing to the purely predictive scope of the SAQM, this means nothing but that some commonsense expectations associated with classical (typically 'non-contextualized') uses of the probability calculus need not apply to *all kinds* of predictive schemes. It certainly does not imply the existence of any sort of nonlocal influences. Manifestations of alleged nonlocality in the quantum-mechanical setting are all notoriously ineffective: 'quantum' correlations cannot be made use of for signalling superluminally or achieving spooky 'transfers' in a controllable manner (quantum teleportation[83], despite its name, is no exception to the rule). Although belief in nonlocality as a pervasive and radically non-classical feature of quantum theory, or as a trait inherent in a 'quantum world', remains predominant, some of the most dedicated explorers of quantum mysteries have voiced a well-informed perplexity:

> "…people may have become too facile in their readiness to blame everything on (or credit everything to) "quantum nonlocality". Nonlocality seems to me to offer "too cheap" a way out of some deep conundra [*sic*] (to appropriate Einstein's remark to Born about Bohm's theory). If you push hard on it you can force "nonlocality" into offering some explanations that strike me as just plain silly….people have been a little too quick in talking themselves into this widely held position[84]."

If the predictive rules of quantum theory can be derived without making any 'ontic' reference to physical systems, it is just as important to realize that the SAQM owes its structure and its actual availability to the existence of a certain class of solutions to the Schrödinger equation, and that this equation can itself be derived as a mere by-product of a principle-based dynamical framework that is regulated by the Q(S)HJE. Fundamental aspects of 'quantum behavior', illustrated e.g. by the two-slit experiment, will not be elucidated as long as baffling invocations of 'self-interference', or the use of half-baked notions like *welcher Weg* (which-way) or even more fashionable quantum information, take precedence over a painstaking study of dynamics at an appropriate protoquantal level (which does not mean Bohm's dynamics). Only after – or if – this is undertaken will our understanding of the quantum measure up to our predictive abilities.

---

[83] Bennett *et al.* 1993.
[84] Mermin 1999.